\documentclass[a4paper,10pt,conference]{IEEEtran}
\IEEEoverridecommandlockouts

\usepackage{amssymb,amsmath,amsthm}
\usepackage{subcaption}
\usepackage{soul,color}
\usepackage{underscore}
\usepackage{graphicx}
\usepackage{booktabs,paralist}
\usepackage[boxed]{algorithm}
\usepackage[noend]{algpseudocode}
\usepackage{cite}
\usepackage{verbatim}
\usepackage{listings}
\usepackage{centernot}
\usepackage{cleveref}
\usepackage{csquotes}
\usepackage{textcomp}
\usepackage{wrapfig}
\newcommand{\mt}[1]{\mathcal{#1}}

\begin{document}
	\title{Formal Synthesis of Monitoring and Detection Systems for Secure CPS Implementations
	\thanks{The authors acknowledge generous support from  Robert Bosch Engineering and Business Solutions Private Limited.}
	}
	
	\author{\IEEEauthorblockN{Ipsita Koley$^{1}$, Saurav Kumar Ghosh$^{1}$, Soumyajit Dey$^{1}$, Debdeep Mukhopadhyay$^{1}$,\\ Amogh Kashyap K N$^{2}$, Sachin Kumar Singh$^{2}$, Lavanya Lokesh$^{2}$,  Jithin Nalu Purakkal$^{2}$, Nishant Sinha$^{2}$}
		\IEEEauthorblockA{\textit{$^{1}$Indian Institute of Technology,Kharagpur, $^{2}$Robert Bosch Engineering and Business Solutions Private Limited}\\
			{\{ipsitakoley, soumyajit, debdeep\}}@iitkgp.ac.in, saurav.kumar.ghosh@cse.iitkgp.ernet.in\\
			{\{Amogh.Kashyap, SachinKumar.Singh, Lokesh.Lavanya, Jithin.NaluPurakkal2,Sinha.Nishant\}}@in.bosch.com
	}}
	\maketitle
	
	\begin{abstract}
		We consider the problem of securing a given control loop implementation of a cyber-physical system (CPS) in the presence of Man-in-the-Middle attacks on data exchange between plant and controller over a compromised network.
		To this end, there exists various detection schemes which provide mathematical guarantees against such attacks for the theoretical control model. However, such  guarantees may not hold for the actual control software implementation. In this article, we propose a formal approach towards synthesizing attack detectors with varying thresholds which can prevent performance degrading stealthy attacks while minimizing false alarms. 
	\end{abstract}
	
	\begin{IEEEkeywords}
		Cyber Physical System, False data injection attack, Formal method, Residue based detector.
	\end{IEEEkeywords}
	
\vspace*{-1mm}
\section{Introduction}
\label{secIntroduction}
Unattended communication among devices in distributed CPS implementations makes new pathways for malicious interference. Given that such systems often need to perform safety critical functionalities with real time deadlines within stringent power, energy requirements, the impact of  attacks on safety-critical CPS may have catastrophic consequences. In the past  decade, many such high profile attacks have been reported spanning a variety of application domains (\cite{karnouskos2011stuxnet,slay2007lessons,khan2016threat,shoukry2013non}). 
It is infeasible to physically secure every packet transmission between CPS components due to limited communication bandwidth as well as lightweight nature of computing nodes. 
This rules out using heavyweight cryptographic encryption techniques (like RSA, AES) along with MACs for securing all intra-vehicular communication \cite{jovanov2018secure}. 
Hence, it makes sense to enhance the security of CPS implementations by using suitable lightweight monitoring primitives considering that an attacker has already breached into the CPS communication infrastructure. 
\par In this work, we focus on residue-based monitoring and detection systems which compute the difference between plant output measurements received through a communication network and the estimates of the same based on earlier measurements and knowledge about system dynamics, raising an alarm if the difference (i.e. the {\em residue}) exceeds a predefined threshold. Since this type of anomaly detector uses the properties of the control system to detect an adversarial action, it does not impose any significant overhead to the system's resource consumption in terms of communication and computation. Although there exists significant literature on residue-based detectors  \cite{liu2011false,sandberg2010security,mo2010false}, none of these works  discusses an effective methodology for synthesizing thresholds given a control system specification. Also existing works consider static thresholds only, i.e. the difference in measurement and estimate is compared with a constant pre-fixed threshold for all closed loop iterations of the system. 
\par As a potential example of targeted  performance  degrading attack, consider the situation when the reference point of a controller changes due to occurrence of some event. 
For such systems, with a comparatively smaller fault injection at the later stage of dynamics (i.e. when nearing the reference), an attacker can prevent the system from reaching the close vicinity of the reference. This brings in interesting trade-offs from the detector design point of view. In a static threshold-based detection scheme,  if the threshold is decided based on the required attack amount at the later phase of settling time, it may be the case that any process or measurement noise induced by environmental disturbance in the system is  considered an attack and a \emph{false alarm} is generated. This implies the False Alarm Rate (FAR) will increase.  If the threshold is decided based on the attacker's effort at the earlier phase of settling time, the attacker can easily bypass the  detection scheme by injecting sufficiently small anomalies whenever the system is very close to the reference and deteriorate the system's performance. This motivates the case for  a variable threshold  based anomaly detection method which may ensure reduced FAR while identifying even small attack  efforts that may lead to potential performance degradation.

 \begin{figure}[H]
		\begin{subfigure}[b]{0.48\columnwidth}\vspace*{-0.1in}
			\includegraphics[width=\linewidth]{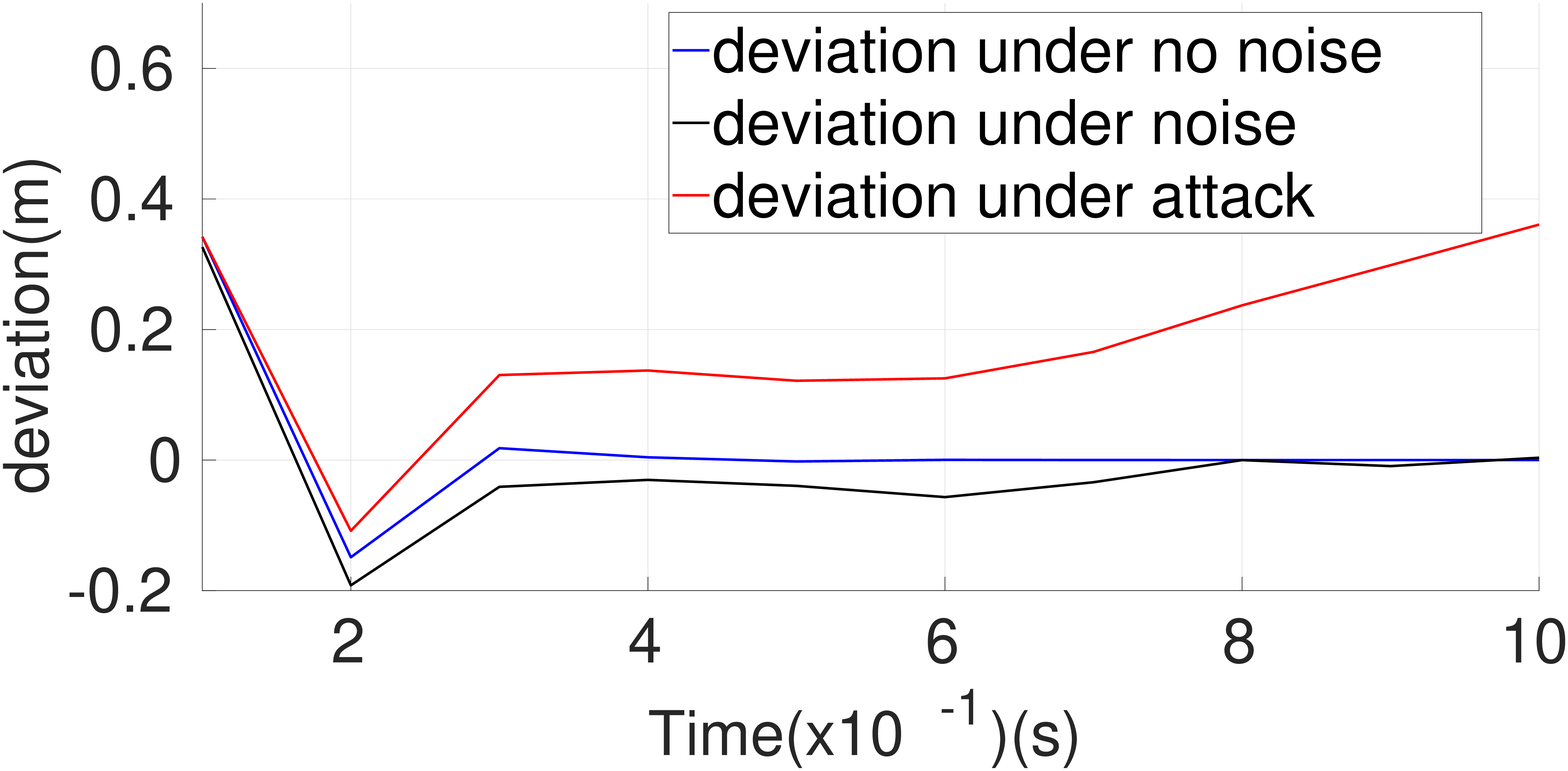}
			\caption{Effect of noise and attack}
		\end{subfigure}
		\begin{subfigure}[b]{0.5\columnwidth}
			\includegraphics[width=\linewidth]{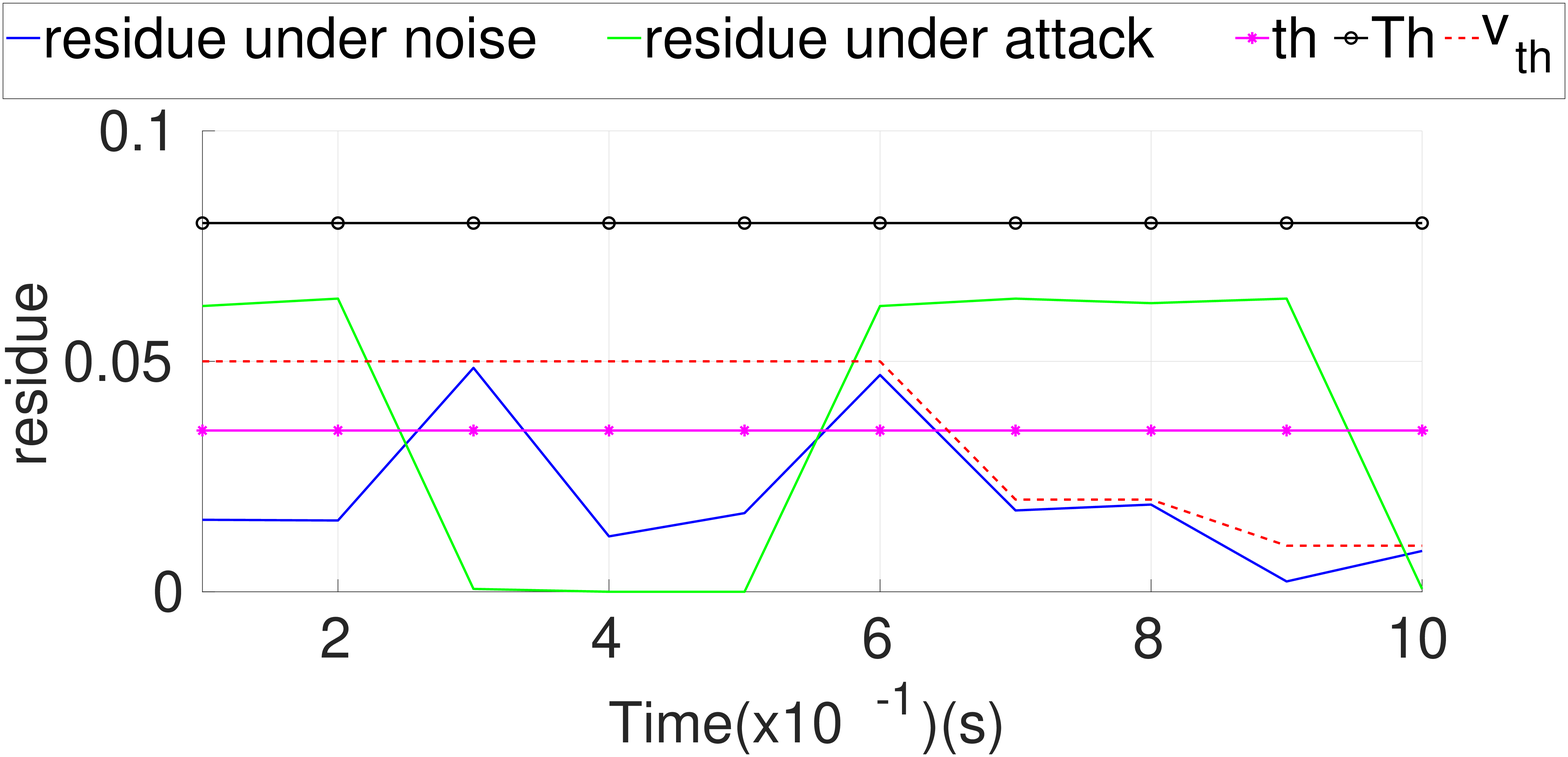}
			\caption{Static vs dynamic threshold}
		\end{subfigure}
		\caption{Trajectory tracking system}
		\vspace*{-0.05in}
		\label{figTrajectoryTracking}
	\end{figure}
\par As a motivational example, we consider a trajectory tracking system  (Fig.~\ref{figTrajectoryTracking}) taken from \cite{kerns2014unmanned}. 
A suitably crafted attack can steer the system towards instability as shown in the same figure. 
In  Fig.~\ref{figTrajectoryTracking}b, we consider three possible residue based detectors, with the smaller threshold $th$, the bigger threshold $Th$ and the variable threshold curve $v_{th}$.
Note that with $th$, the detector considers even the  harmless noise as an attack, as shown  (Fig.~\ref{figTrajectoryTracking}b). On the other hand,  with $Th$, the actual attack could easily bypass the detector. However using the variable  threshold curve $v_{th}$ (dotted red line in Fig.~\ref{figTrajectoryTracking}b), the  attack does not remain stealthy while harmless noise is allowed to pass reducing the FAR. 
\par In this article, 
we propose a formal approach for synthesizing residue based attack detectors with variable thresholds for CPS implementations  that can prevent stealthy attacks. These detectors are also guaranteed to have  smaller FAR w.r.t. provably safe static threshold based detector options.

\section{System Model and Problem Formulation}
\label{secSystemModel}
Consider a discrete linear time invariant (LTI) plant model $\mt{S}$ given as, $x_{k+1} = Ax_{k} + Bu_{k} + w_{k}, y_{k} = Cx_{k} + Du_{k} + v_{k}$, where $x_{k}\in\mathbb{R}^{n}$, $y_{k}\in\mathbb{R}^{m}$, $w_{k}\in\mathbb{R}^{n}\sim \mathcal{N}(0, Q)$ and $v_{k}\in\mathbb{R}^{m}\sim\mathcal{N}(0,R)$ represent system state variables, sensor measurements of plant, zero mean Gaussian process and measurement noise at $k^{th}$ sampling instance respectively. Also $A,\, B$ and $C$ are transition matrix, input map and output map for the plant model respectively. To estimate the system states $\hat{x}_{k}$ from the observed ones, a Kalman filter based observer is deployed, given by, $z_{k} = y_{k} - C\hat{x}_{k},\, \hat{x}_{k+1} = A\hat{x}_{k} + Bu_{k} + Lz_{k}$, where residue $z_{k}=y_k-C\hat{x}_k$ is the difference between measured and estimated output at $k^{th}$ time instance, and $L$ is Kalman gain.
The controller output $u_{k}$ is computed as, $u_{k} = -K\hat{x}_{k}$.
In this paper we contemplate false data injection  based attack scenario in which the attacker falsifies the sensor measurements by injecting $a_{k} \in \mathbb{R}^{m}$ to sensor output $y_{k}$ at $k^{th}$ sampling instance.
The resulting altered sensor measurements $\tilde{y}_{k} = y_{k} + a_{k}$ are fed to the estimator   which in turn affects the control input calculation. Due to this, the  closed loop dynamics deviates from the expected behavior. For  this system description, we consider a threshold based detection scheme such that the detector will raise an alarm whenever $\lVert z_{k}\rVert \geq Th[k]$ where $Th[k]$ is threshold at $k^{th}$ sampling instance. We say an attack is $stealthy$ if some given safety or performance criteria of the system is violated by the attacker while $\lVert z_{k} \rVert$ remains below $Th[k]$ for all $k$. 
\\
\textbf{Formal Problem Statement:} Consider the plant model $\mathcal{S}$ as discussed earlier, a controller implemented as a software program $\mathcal{C}$ running on an ECU, and a safety or performance criteria $pfc$. The objective of $\mathcal{C}$ is to satisfy $pfc$ withing some $(j + T)^{th}$ samples starting from any sampling instance $j$. An $l$ length threshold specification $Th$ is represented by a vector $\in \mathbb{R}^l$. Threshold is said to be \emph{static} if $Th[i]$ is same for all $i$, else it is \emph{variable}. We formally define the threshold synthesis problem as follows.\\ {\em Given $\langle \mathcal{S}, \mathcal{C}, pfc \rangle$, what would be an optimal threshold specification $Th$ such that any stealthy attack is guaranteed to be detected as well as FAR is minimized ?}
\section{Threshold Synthesis and Methodology}
\label{secThresholdSynthesis}
As a first step in our approach, we propose Algorithm~\ref{algAttackSyn} which formally checks the implementation $\mt{C}$ and identifies whether there exists any possible attack vector that can violate the  target properties of the system. Given $\mathcal{S},\,\mathcal{C}$, let  $x_{des}$ be the reference point for the system and the target property $pfc$ is to reach $x_{T} \in \{x_{des} + \epsilon\}$ for some $\epsilon \in \mathbb{R}$ within a finite number of iterations, say $T$ starting from any initial state $x_1\in \mt{V}\subseteq \mathbb{R}^n$. An  attacker would  want to achieve $x_{T} \centernot \in \{x_{des} + \epsilon\}$ after $T$ such  closed loop iterations. Some CPS implementations often incorporate certain monitoring constraints in addition to residue based attack detectors to check the sanity of the sensor measurements. Such constraints are captured using suitable predicates denoted as $mdc$. Algorithm~\ref{algAttackSyn} takes as input $mdc$, $pfc$, a threshold vector $Th$ and a finite duration $T$ allotted for achieving $pfc$. The variable $a_{k}$ signifying false data is assigned a value nondeterministically (Line 4) and is added with measurements subsequently. We say that an attack is {\em stealthy} but successful when predicates $\lVert z_{k}\rVert \leq Th[k]$ and $mdc$ are satisfied, but $pfc$ is violated. This is modeled by the assertion $\mathbb{A}$ in Line $9$. $\mathbb{A}$ is given as input to an SMT tool with the assert clause. It returns a successful attack vector $\mathcal{A}$ if the assertion $\mathbb{A}$ is satisfied (Line $11$). Otherwise, it returns NULL (Line $13$) which  guarantees that no attack vector exists that remains stealthy over $T$ iterations and violates the performance criteria $pfc$ of the system.

\begin{algorithm}\scriptsize
		\caption{Attack vector synthesis}
		\label{algAttackSyn}
		\begin{algorithmic}[1]
			\Require{Control property $pfc$, existing monitoring constraint $mdc$, computed threshold vector $Th$, attack duration $T$}
			\Ensure{Attack vector $\mathcal{A}$(if it exists, otherwise NULL)}
			\Function{AttVecSyn}{$Th$, $pfc$, $mdc$, $T$}
			\State $x_{1} \gets \mathcal{V}$; $\hat{x}_{1}\gets0$;  $u_{1} \gets 0$; \Comment{Initialization}
			\For{$k=1$ to $T$}
			\State $a_{k} \gets nondeterministic\_choice$;
			\State $y_{k} \gets Cx_{k} + Du_{k} + a_{k}$; $\hat{y}_{k} \gets C\hat{x}_{k} + Du_{k}$;
			\State $z_{k} \gets y_{k} - \hat{y}_{k}$;
			\State $x_{k+1} \gets Ax_{k} + Bu_{k}$; $\hat{x}_{k+1} \gets A\hat{x}_{k} + Bu_{k} + Lz_{k}$;
			\State $u_{k+1} \gets -K\hat{x}_{k+1}$;
			\EndFor
			\State $\mathbb{A}\gets$\textbf{assert}(($\forall Th[p]\in Th, \lVert z_{p}\rVert < Th[p]$ \&\& $mdc$) $\centernot\to$ $pfc$);
			\If {$\mathbb{A}$ is {\em valid}}
			\State \Return $\mathcal{A} \gets
			\left[a_{1} \cdots a_{T} \right]$;
			\Else
			\State\Return NULL;
			\EndIf			
			\EndFunction
		\end{algorithmic}
	\end{algorithm}	
We now propose a methodology in Algorithm~\ref{algSporadicThresholdSyn} to synthesize a monotonically decreasing vector of thresholds to provably secure a given CPS against attacks. Given  the state space of possible $l$-length variable threshold functions ($l\in\mathbb{N}$), we  formulate heuristic  approaches guided by our hypothesis of monotonically decreasing thresholds. To verify whether existing monitoring constraint (if any) suffices to detect any stealthy attack we generate an attack vector without any threshold based detector (Line $2-3$) using Algorithm~\ref{algAttackSyn}. If any attack vector is retrieved, we make a greedy choice and select the sampling instance $i$ where maximum residue is generated due to this attack (Line $4$) as a pivot point. A threshold at $i$ is introduced to thwart the current attack (Line $5$). With this new threshold we call Algorithm~\ref{algAttackSyn} (Line $6$) to check if any attack can bypass this detector. If found, we now search for new thresholds to be added to $Th$ to stop this new attack in the following manner.\\
\emph{Case $1a$} [Line $9-11$]: For any of the existing thresholds $Th[p] \in Th$, we try to find out whether the current attack has produced any residue $\lVert z_k  \rVert \geq Th[p]$ before the  $p^{th}$ instance, i.e. $k\leq p$. If any such $z_k$ exists, we consider the maximum of them and include it to $Th$ while ensuring monotonicity (Line $9-10$). If we get such a new threshold $Th[i]$ that keeps the monotonic decreasing order in $Th$ intact, we stop searching (Line $11$). Otherwise, we consider \emph{Case} $1b$.\\ \vspace*{-2mm}
\begin{algorithm}[H]\scriptsize
	\caption{Pivot Based Threshold Synthesis}
	\label{algSporadicThresholdSyn}
	\begin{algorithmic}[1]
		\Require{Performance criteria $pfc$, existing monitoring constraint $mdc$, number of sampling instances $T$ required by the controller to attain $pfc$}
		\Ensure{Threshold Vector $Th$ = \{ $Th[i]$: threshold required at $i^{th}$ sampling instance to thwart false data injection attack\}} of length $T$
		\Function{PivotBasedThresholdSyn}{$pfc$, $mdc$, $T$}
		\State $Th \gets NULL$; \Comment{Initialization}
		\If{\Call{AttVecSyn}{$Th$, $pfc$, $mdc$, $T$}}
		\State $\exists i\in[1,T]\;, \forall\;k\in[1,T], \;i\neq k\wedge \lVert z_{i}\rVert \geq \lVert z_{k}\rVert$
		\State $Th[i] \gets \lVert z_{i}\rVert$;
		\EndIf
		\While {\Call{AttVecSyn}{$Th$, $pfc$, $mdc$, $T$}}	
		\State $found \gets false$;
		\For{each $p\in[1,T]$ s.t. $Th[p] \neq 0$} \Comment{New threshold addition}
		\If{$\exists i\;\lVert z_i\rVert\gets max(\forall k\in[1,p-1]\;\lVert z_k\rVert\wedge\lVert z_k\rVert\geq Th[p])$}
		\State $Th[i] \gets min(\forall k\in[1,i-1]\;Th[k]\neq0,\lVert z_i\rVert)$;
		\State $found \gets true$;\;{\bf break};
		\EndIf
		\If{$\exists i\;\lVert z_i\rVert\gets max(\forall k\in[p+1,T]\;\lVert z_k\rVert)$}
		\If{$\forall k\in[i+1,T]\;\lVert z_i\rVert\geq Th[k]$}
		\State $Th[i] \gets min(\forall k\in[1,i-1]\;Th[k]\neq0, \lVert z_i\rVert)$;
		\State $found \gets true$;\;{\bf break};
		\EndIf
		\EndIf
		\EndFor
		\If{$found = false$} \Comment{Threshold reduction step}
		\State $\exists i\in[1,T]\; \forall k\in[1,T] \;i\neq k\wedge Th[i]\neq 0 \wedge Th[k]\neq 0 \wedge ((Th[i]-\lVert z_{i}\rVert)\leq(Th[k]-\lVert z_{k}\rVert))$;
		\State $Th[i] \gets\lVert z_{i}\rVert$;
		\For {all $k\in[i+1,T]$}
		\If {$Th[k]>Th[i]$}
		\State $Th[k]\gets Th[i]$
		\EndIf
		\EndFor
		\EndIf
		\EndWhile
		\State \Return $Th$
		\EndFunction
	\end{algorithmic}
\end{algorithm}
\emph{Case $1b$} [Line $12-15$]: For any of the thresholds $Th[p] \in Th$, we try to find out whether the current attack has produced any residue $\lVert  z_k \rVert \geq Th[j]$ for all $j \in [k+1, T]$ where $k > p$. In that case, we consider the maximum of them (Line $12$) and include it to $Th$ while ensuring monotonicity (Line $13-14$). Otherwise, one or more existing thresholds in $Th$ need to be reduced to detect the current attack (\emph{Case $1c$}).\\
\emph{Case $1c$} [Line $16-21$]: We choose the candidate threshold $Th[i]$ from $Th$ which can be reduced with minimum  effort i.e. the minimum difference between the current threshold value $Th[i]$ and the residue  $\lVert z_{i}\rVert$ generated by the attack (Line $17$). For that $i$, we set $Th[i]=\lVert z_i\rVert$ (Line $18$) and adjust subsequent thresholds in order to ensure monotonicity (Line $19-21$).\\
Once a new threshold is introduced or existing thresholds are modified to detect the current attack, we call  Algorithm~\ref{algAttackSyn} (Line $6$) with the modified $Th$. If it returns NULL, it is ensured that the latest $Th$ is enough to thwart any stealthy attack. If not, we repeat the process with Case $1a$, $1b$ or $1c$ with the newly generated attack vector. While Algorithm~\ref{algSporadicThresholdSyn} can be used to synthesize monotonically decreasing thresholds, it can take a long time to converge. Hence we propose Algorithm~\ref{algStepThresholdSyn} which also starts with generating an attack vector without considering any threshold using Algorithm~\ref{algAttackSyn} and finds the sampling instance $i$ at which maximum residue is generated (Line $3-4$). Considering a staircase approximation of the target variable threshold vector, we maintain the vector $Steps$ to keep track of the heights of the step edges of the staircase where $Steps[k]$ denotes  height of the $k^{th}$ step. In this algorithm, a step captures a subsequence of consecutive constant thresholds. First step of staircase is created by setting $\forall 1\leq j\leq i,\,Th[j] = Steps[i]$ where $Steps[i] = \lVert z_{i}\rVert$ (Line $5-6$). With this new threshold vector $Th$, we call Algorithm~\ref{algAttackSyn} to check if any attack can bypass this detector. If yes, we generate new threshold steps in the following ways.
\vspace*{-1mm}
\begin{algorithm}[H]\scriptsize
	\caption{Step-wise Threshold Synthesis}
	\label{algStepThresholdSyn}
	\begin{algorithmic}[1]
		\Require{Performance criteria $pfc$, existing monitoring constraint $mdc$, number of sampling instances $T$ required by the controller to attain $pfc$}
		\Ensure{Threshold Vector $Th$ = \{ $Th[i]$: threshold required at $i^{th}$ sampling instance to thwart false data injection attack\}}
		\Function{StepWiseThresholdSyn}{$pfc$, $mdc$, $T$}
		\State $Th \gets NULL; Steps \gets NULL$; \Comment{Initialization}
		\If{\Call{AttVecSyn}{$Th$, $pfc$, $mdc$, $T$}}
		\State $\exists i\in[1,T]\; \forall\;k\in[1,T] \;i\neq k\wedge \lVert z_{i}\rVert \geq \lVert z_{k}\rVert$
		\State $Steps[i]\gets\lVert z_{i}\rVert$;
		\State $\forall 1\leq j\leq i,Th[j] \gets Steps[i]$;$k\gets i$;
		\EndIf
		\While{\Call{AttVecSyn}{$Th$, $pfc$, $mdc$, $T$} $\wedge \; k\neq T$}\Comment{Initial steps formation}
		\State $\exists i\; i<T \wedge Th[i]\neq0 \wedge Th[i+1]=0$;
		\State $\exists k\; i<k<T\; \forall j\;i<j<T\;\wedge j\neq k\;\wedge \lVert z_j\rVert\leq \lVert z_k\rVert \leq Th[i]$;
		\State $Steps[k]\gets \lVert z_{k}\rVert$;
		\State $\forall i< j\leq k \; Th[j] \gets Steps[k]$;
		\EndWhile	
		\While{$\mathcal{A} \gets$ \Call{AttVecSyn}{$Th$, $pfc$, $mdc$, $T$}} \Comment{Reducing height of steps}
		\State $k \gets$\Call{MinAreaRectangle}{$\mathcal{A}$,$Steps$,$T$};
		\State $\exists p\;k<p<T\; \forall q \;k<q<T \;\wedge p\neq q \wedge Steps[p] \leq \lVert z_k\rVert \wedge Steps[q] \leq \lVert z_k\rVert \wedge Steps[p]\geq Steps[q]$;
		\State $Steps[k]\gets Th[k]$; $Steps[p]\gets\lVert z_{k}\rVert$;
		\State $\forall i\;k< i\leq p,Th[i] \gets Steps[p]$;
		\EndWhile
		\State \Return $Th$
		\EndFunction
		\Function{MinAreaRectangle}{$\mathcal{A}$,$Steps^\prime$,$T$}
		\State $MinArea\gets\infty;MinAreaPositon\gets NULL;$
		\For{$i=1$ to $T$}
		\State $Area_{i} \gets0;$
		\While {$\exists j\;i<j<T\; \forall k \;i<k<T \;\wedge j\neq k \wedge  Steps^\prime[j]\geq Steps^\prime[k] \wedge Steps^\prime[j] > \lVert z_i\rVert$};
		\State $Area_{i} \gets Area_i + (Steps^\prime[j]-\lVert z_{i}\rVert)\times(j-i)$;
		\State $Steps^\prime[j]\gets NULL;$
		\EndWhile
		\If{$Area_i<MinArea$}
		\State $MinArea\gets Area; MinAreaPosition\gets i;$
		\EndIf
		\EndFor
		\State \Return $MinAreaPosition$;
		\EndFunction
	\end{algorithmic}
\end{algorithm}
\emph{Case $2a$} [Line $7-11$]: Let $i$ be the last step with non-zero threshold (Line $8$). To generate a new step after $i$, we find out the sampling instance $k$ at which the maximum residue is generated by the current attack vector such that $k>i$. The record of new step edge $Steps[k] = \lVert z_{k}\rVert$ is added to $Steps$ vector (Line $10$) and the new step is enforced by setting  $\forall\;j\in (i, k],\; Th[j] = Steps[k]$ (Line $11$). If the last step edge is at $i=T$ or no stealthy attack can be found that bypasses the current threshold steps, we proceed to \emph{Case $2b$} to build new steps by fine-graining the existing ones.\\
\emph{Case $2b$} [Line $12-17$]: In this case, we have two possibilities. If no attack vector exists  (Line 12), then the algorithm terminates. If any attack is found that bypasses the current detector threshold $Th$, heights of the existing steps need to be reduced. Instead of diminishing the height of an entire step, we break a portion or the whole step whichever involves minimum effort i.e. the minimum area from under the threshold curve that can be removed to detect the current attack. The function \Call{MinAreaRectangle}{} (Line $18-27$) computes such minimum area ensuring  both staircase like structure and monotonic decreasing property.

\section{Case Study and Observations}
\label{secCaseStudy}
We demonstrate the efficacy of our approach using a Vehicle Stability Controller (VSC) case study. 
The VSC system receives data from four wheel speed sensors (WSS), lateral acceleration (Ay), longitudinal acceleration (Ax), yaw rate sensor (Yrs) and steering angle sensor (SaS). Generated actuator command is sent to the hydraulic unit of a vehicle. Wheel speed sensors are hardwired between the wheels and the controller unit. However, data from Ay, Ax, Yrs, SaS, along with actuator signal, are transferred through CAN bus and is  considered vulnerable to attack. In this work, we use VSC model of \cite{aoki2005experimental}.
Sampling period is considered as $T_{s} = 40 ms$. Relevant variables are 
taken from \cite{zheng2006controller}. 
We consider an attack model where the attacker forges output of both $Yrs$ and $Ay$ sensors. However, most   modern automobiles have monitoring systems already in place to detect any abnormal behavior of VSC. We consider one such monitoring system which performs the following checks for all measurements: 1) \emph{Range and gradient based monitors} check if range and gradient of yaw rate $\gamma$ and lateral acceleration $a_y$ are within permissible limit; 2) \emph{Relation based monitor} checks if difference between measured yaw rate $\gamma$ from Yrs and estimated yaw rate $\gamma_{est}$ from Ay is less than $allowedDiff$.
An immediate violation of both the schemes does not raise an alarm. It waits for certain duration, called \emph{dead zone}. Continuous violation during the dead zone causes the monitoring system to raise an alarm. The $allowedDiff$, range of $\gamma$, gradient of $\gamma$, range of $a_y$ and gradient of $a_y$ are considered $0.035$ rad/s, $0.2$ rad/s, $0.175$ rad/s$^{2}$, $15$ m/s$^{2}$ and $2$ m/s$^{3}$ respectively. The dead zone is considered to be $300$ ms i.e. $\lfloor \frac{300}{T_s} \rfloor = 7$ samples. We define $pfc$ of the system  as: yaw rate must reach within $80$\% of desired value within $50$ sampling instances.
\begin{figure}[H]
	\centering
	\begin{subfigure}[b]{0.42\columnwidth}
		\includegraphics[width=\linewidth, height=2.2cm]{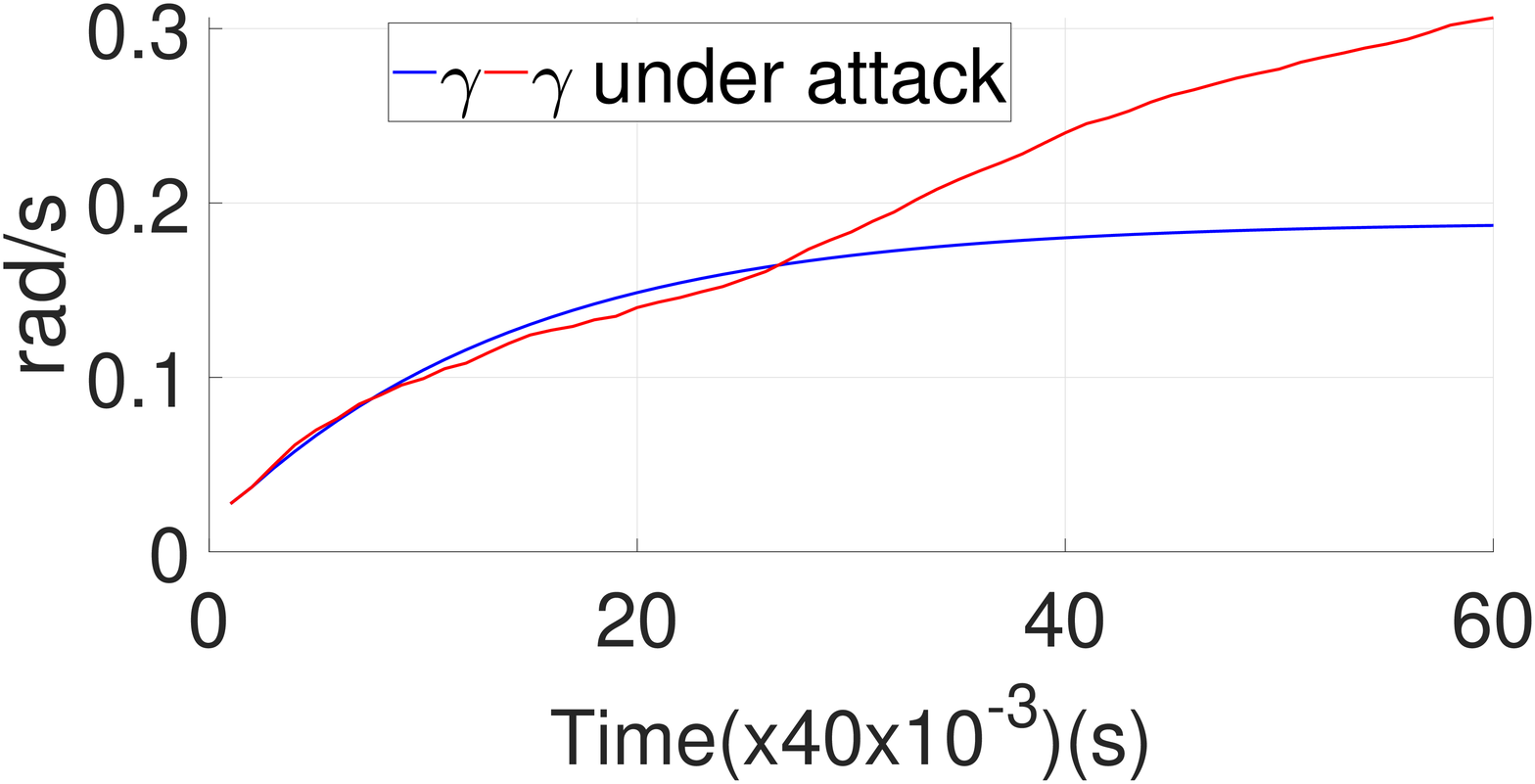}
		\caption{Plant state $\gamma$}
	\end{subfigure}
	\hspace{-2mm}
	\begin{subfigure}[b]{0.55\columnwidth}
		\includegraphics[width=\linewidth]{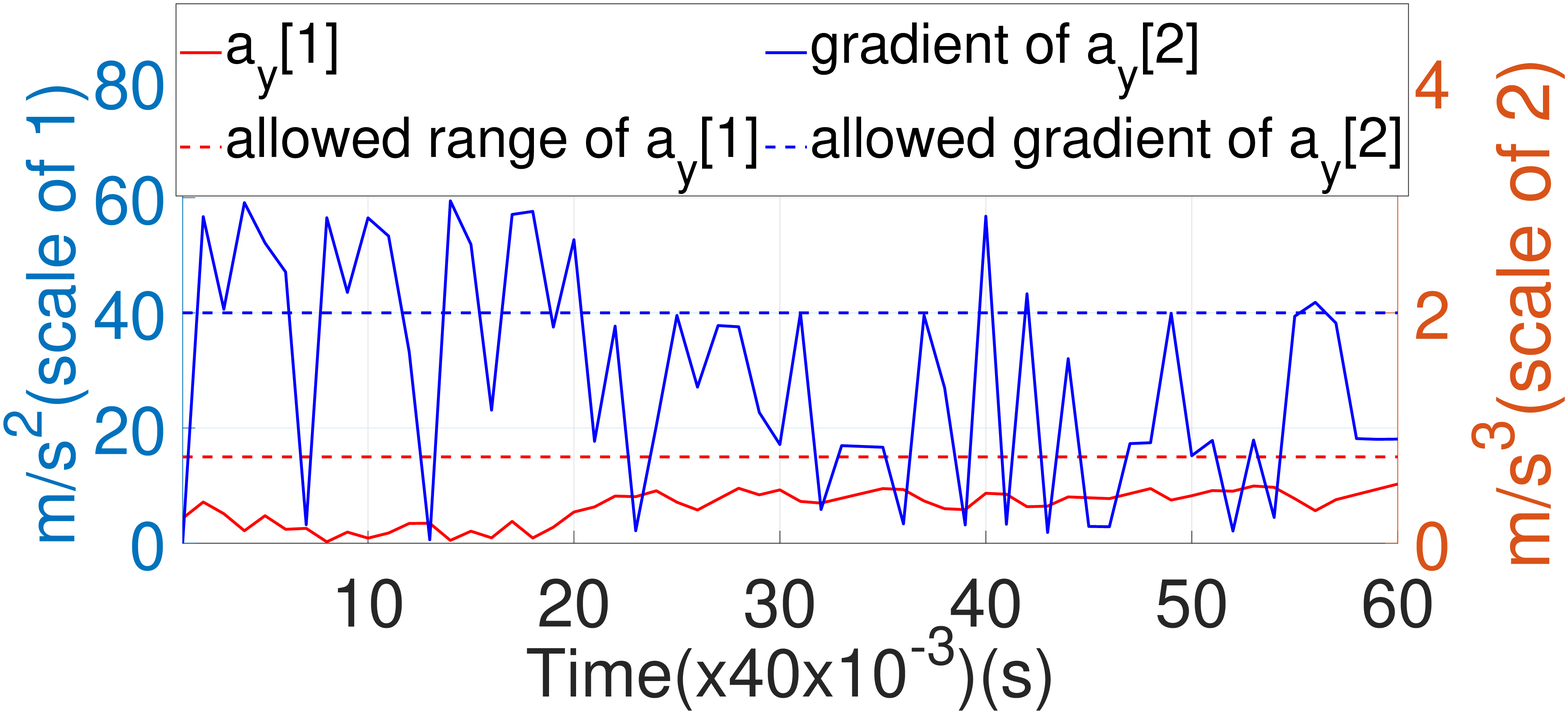}
		\caption{Monitoring on $a_y$}
	\end{subfigure}
	\begin{subfigure}[b]{0.8\columnwidth}
		\includegraphics[width=\linewidth]{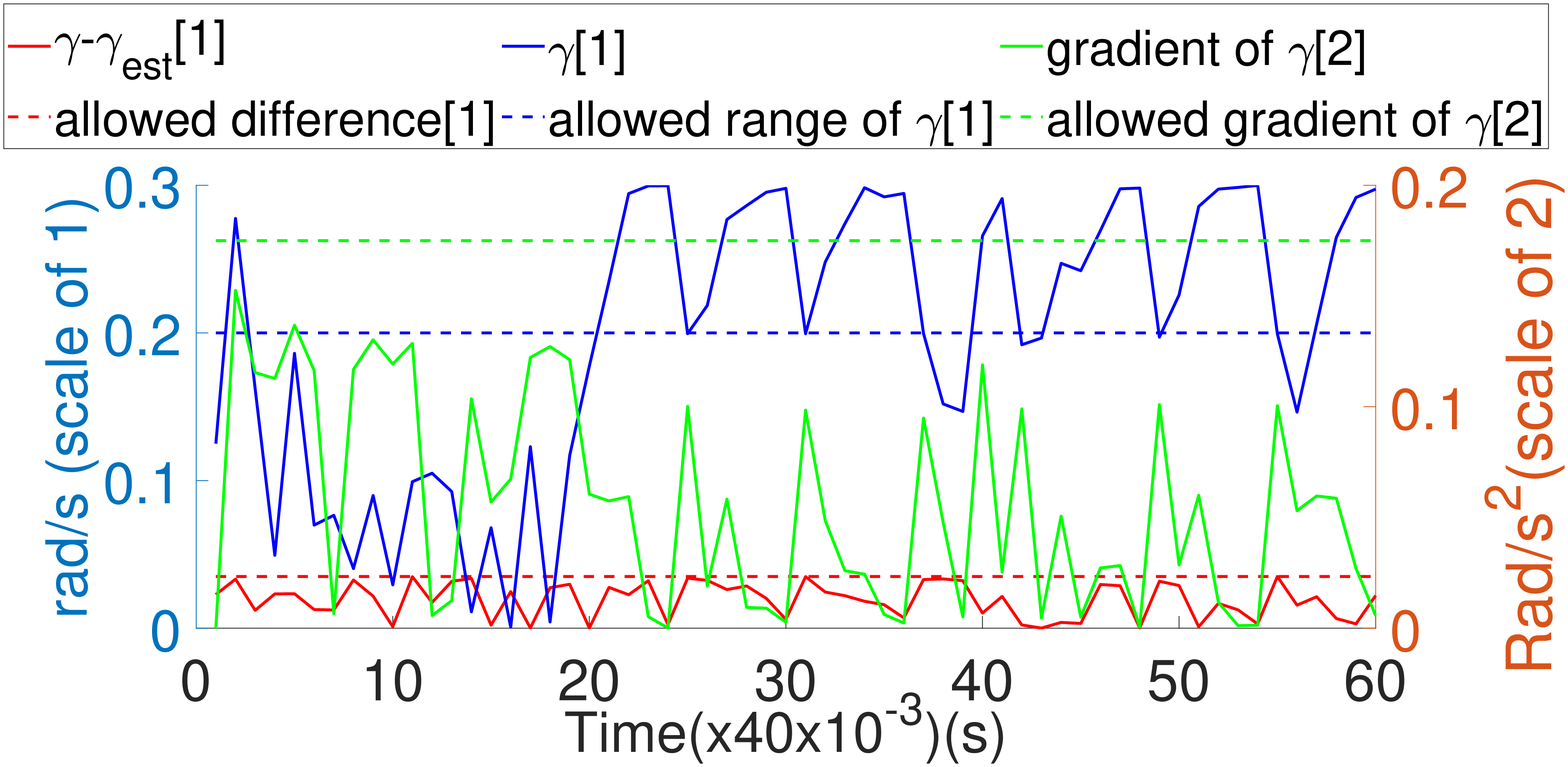}
		\caption{Monitoring on $\gamma$}
	\end{subfigure}
	\caption{Attack demonstration on VSC}
	\label{figAttackOnVSC}
\end{figure}
To verify whether this apparently efficient monitoring system can be bypassed by an attacker, we formulate an SMT problem  in Algorithm~\ref{algAttackSyn}. We model all monitors as predicate $mdc$ in Algorithm~\ref{algAttackSyn}. We include $pfc$ and $mdc$ in the assertion clause $\mathbb{A}$ (Line $9$ of Algorithm~\ref{algAttackSyn}) and use the popular SMT solver Z3\cite{de2008z3}. The output array $\mathcal{A}$, in Algorithm~\ref{algAttackSyn}, if nonempty, reports attack vectors for the system. The effect of one such synthesized vector for the VSC system is demonstrated in Fig.~\ref{figAttackOnVSC}a. The attack bypasses the existing monitoring schemes (Fig.~\ref{figAttackOnVSC}b,c). For mitigating these vulnerabilities, we synthesize suitable  residue based detectors using our methods.


\begin{wrapfigure}{r}{0.5\columnwidth}
	\centering
	\includegraphics[scale=0.1,trim={2cm 0 2cm 0},clip]{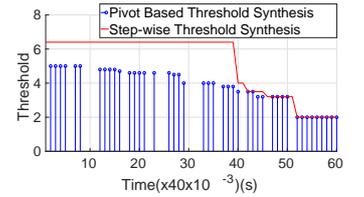}
	\caption{Output of variable threshold synthesis Algorithms}
	\label{figThresholdPlot}
\end{wrapfigure}
With $pfc$, $mdc$ of VSC and $T$ as input, we execute Algorithms~\ref{algSporadicThresholdSyn} and \ref{algStepThresholdSyn} with a timeout of $12$ hours for each SMT call. Based on the greedy choices made during simulation, Algorithm~\ref{algSporadicThresholdSyn} terminates in the $56^{th}$ round while Algorithm~\ref{algStepThresholdSyn} terminates much faster, in the $37^{th}$ round. The final threshold sets computed by both algorithms are presented in  Fig.~\ref{figThresholdPlot}. For comparison purpose, we also synthesize a static threshold based detector for VSC. We generate $1000$ random measurement noise vectors of bounded length with each value sampled from a suitably small range such that $pfc$ is maintained. Among these, we discard the noise vectors that are detected by $mdc$. From the remaining, we compute false alarm rate of the three threshold based detectors as: a) $61.5\%$ for Algorithm~\ref{algSporadicThresholdSyn}, b) $45.6 \%$ for Algorithm~\ref{algStepThresholdSyn}, and c) $98.9 \%$ for static threshold based detector. We can see that both our proposed algorithms outperform static threshold based detector in terms of FAR. 
\section{Conclusion}
\label{secConclusion}
In the present work, we provide a synthesis mechanism for variable threshold based detectors in the context of securing CPS implementations. Our approach, based on formal techniques, can provide provable guarantees for an actual controller implementation instead of probabilistic guarantees as is  standard for mathematical control models. In future, we would like to perform more exhaustive experimental as well as analytical evaluation of our proposed techniques.  
	
	\bibliographystyle{IEEEtran}
	{\scriptsize
	\bibliography{IEEEabrv,date2019}}
	
\end{document}